\documentclass[twocolumn, amssymb, nobibnotes, aps, pra, superscriptaddress, 10pt]{revtex4-2}
\usepackage{bbm}
\pdfoutput=1
\usepackage[utf8]{inputenc}
\usepackage[english]{babel}
\usepackage[T1]{fontenc}
\usepackage{bm}
\usepackage{mathtools}
\usepackage{amssymb}
\usepackage{amsthm}
\usepackage{physics}
\usepackage{braket}
\usepackage[colorlinks=True, linkcolor= red, citecolor= blue]{hyperref}
\usepackage{nameref}
\usepackage{enumitem}
\usepackage{xcolor}

\def\Var#1{\left(\Delta{#1}\right)^2}
\def\Ex#1{\left\langle {#1} \right\rangle}
\def\eexp#1{e^{#1}}

\newcommand{\sig}{b}
\newcommand{\est}{\sig_{\text{est}}}

\begin{document}


\title{Entanglement Enhanced Sensing with Qubits affected by non-Markovian Dephasing}

\author{Noah Kaufmann}
\affiliation{Center for Hybrid Quantum Networks (Hy-Q), The Niels Bohr Institute, University of Copenhagen,  Jagtvej 155A, Copenhagen DK-2200, Denmark}

\author{Kasper H. Nielsen}
\affiliation{Center for Hybrid Quantum Networks (Hy-Q), The Niels Bohr Institute, University of Copenhagen,  Jagtvej 155A, Copenhagen DK-2200, Denmark}
\affiliation{NNF Quantum Computing Programme, The Niels Bohr Institute, University of Copenhagen,  Blegdamsvej 17, Copenhagen DK-2100, Denmark}

\author{Eva M. González-Ruiz}
\affiliation{Center for Hybrid Quantum Networks (Hy-Q), The Niels Bohr Institute, University of Copenhagen,  Jagtvej 155A, Copenhagen DK-2200, Denmark}

\author{Anders S. Sørensen}
\affiliation{Center for Hybrid Quantum Networks (Hy-Q), The Niels Bohr Institute, University of Copenhagen,  Jagtvej 155A, Copenhagen DK-2200, Denmark}

\begin{abstract}
    Entanglement has been proposed as a means to improve the sensitivity of sensing weak signals. While the degree of this quantum advantage is well understood in noiseless settings, the situation is more complex under realistic conditions, where the system is subject to decoherence. In this case, the enhancement depends on the specific noise characteristics. Previous treatments of colored noise typically assume that the noise is uncorrelated between successive experiments. Here, we consider the scenario in which the noise exhibits correlations across multiple shots. We derive a simple fundamental limit to the sensitivity based on the fact that the sensitivity cannot be better than the signal-to-noise ratio seen by the probe. Focusing on Ramsey spectroscopy with probes affected by pure classical dephasing, we show that, for suitable spatial and temporal noise correlations, entangled probes achieve a better scaling of the sensitivity with the number of probes than separable states. This demonstrates that entanglement can provide a substantial improvement for Ramsey spectroscopy subject to correlated noise.
\end{abstract}

\maketitle


\section{Introduction}
High-precision measurements are central to exploring new physics~\cite{Gilmore2021}. An improved understanding of physical processes, in particular in atomic, molecular, and optical systems, has led to the development of highly sensitive sensors~\cite{Ye2024}. Such devices enable the measurement of even minuscule effects, thereby opening a window into previously unobserved phenomena and enabling an improved understanding of nature. To further push the frontier in sensitivity, quantum entanglement has been suggested~\cite{Caves1981} and experimentally demonstrated~\cite{LIGO2019, VIRGO2019}, as a next resource for enhancing measurement precision.

The uncertainty in estimating a parameter from $N$ independent probes scales as $1/\sqrt{N}$, a consequence of the central limit theorem~\cite{Giovannetti2006}. Quantum enhanced sensing aims to surpass this so-called standard quantum limit by taking advantage of entangled probes~\cite{Giovannetti2004}. The ultimate bound imposed by quantum mechanics, called the Heisenberg limit, scales as $1/N$~\cite{Bollinger1996, Ou1997} and is achievable in noiseless settings~\cite{Yurke1986, Bollinger1996}. In realistic, noisy settings, however, entanglement does not necessarily enable improved scaling~\cite{Huelga1997, Escher2011}, and whether the advantage persists depends on the precise noise characteristics. 

A widely studied paradigm for exploring entanglement-enhanced sensing employs various types of Ramsey spectroscopy~\cite{Degen2017}. Typically, the sensors are two-level systems whose energy splitting depends on the amplitude of an external field, enabling its estimation. When these sensing qubits are prepared in a superposition of energy eigenstates and exposed to the external field, they accumulate a relative phase that depends on the field strength. By subsequently measuring that phase, one can estimate the external field strength. The estimation uncertainty arises both from the probabilistic measurement outcomes and from noise affecting the phase evolution. Given fixed resources, namely the total interrogation time $T$ and the number of two-level systems $N$, and assuming no bias, the objective is to minimize the estimation uncertainty of the external field by optimizing the initial state preparation, the measurement procedure, and the number of shots of the experiment $L$ that are conducted within the total time $T$. In particular, we are interested in the asymptotic estimation uncertainty scaling, which describes the limit for large $N$ and $T$~\cite{Haase2016}.

For separable resources and in the absence of noise, it is optimal to prepare each of the qubits in an even superposition of the energy eigenstates, leading to the standard quantum limit $1/\sqrt{N}$~\cite{Giovannetti2006}. The first proposal for entanglement-enhanced sensing in this setting involved the preparation of spin-squeezed states~\cite{Wineland1992, Wineland1994}. Later work showed that, under noiseless conditions, maximally entangled GHZ states are optimal and achieve the Heisenberg scaling of $1/N$~\cite{Bollinger1996}. However, when qubits are affected by independent Markovian dephasing, estimation protocols using GHZ states lose their advantage over separable strategies~\cite{Huelga1997}, while partially entangled states, such as spin-squeezed states, offer an improvement by a constant factor~\cite{Huelga1997, Orgikh2001}. Indeed, it has been proven that surpassing a scaling of $1/\sqrt{N}$ is not possible in the presence of independent Markovian dephasing~\cite{Escher2011}. For non-Markovian noise, in contrast, it was argued that an estimation uncertainty scaling of $1/N^{3/4}$ is achievable with GHZ states~\cite{Matsuzaki2011, Chin2012} and that this scaling, known as the Zeno limit, is optimal~\cite{Macieszczak2015}. This scaling was shown to be a general property of dephasing noise that is correlated at short times~\cite{Smirne2016, Haase2018}, and has been demonstrated experimentally~\cite{Long2022}. 

However, these results for colored noise were derived under the assumption of statistically independent experimental shots~\cite{Matsuzaki2011, Chin2012, Macieszczak2015, Smirne2016, Haase2018, Riberi2022, Riberi2023, Yang2024_nature, Long2022}. Although often not stated explicitly, this assumption enters through the standard application of the quantum Cramér–Rao bound ~\cite{Braunstein1994}, where the estimation uncertainty for $L$ experimental shots is obtained by dividing the single-shot uncertainty by the square root of the number of shots, $\sqrt{L}$. This treatment is not valid in the case of correlations between successive experimental shots. For temporally correlated noise, this treatment therefore assumes that the noise is effectively reset after every Ramsey sequence~\cite{Macieszczak2015, Yang2023, OConnor2024}. This framework is therefore also referred to as fresh-environment approach~\cite{Das2026}.  While such an assumption may be justified for some experimental scenarios, e.g., cold atom experiments~\cite{Ockeloen2013}, which have a limited interrogation time before the atoms are lost, it is generally not justified when the temporal correlations extend across multiple experimental shots.

Generally, noise correlations across experimental shots become relevant in experimental scenarios where the noise has long-term correlations. For instance, this applies to any system with a Ramsey decoherence time $T_2^*$ shorter than the coherence time $T_2^{\text{echo}}$ measured using a spin-echo sequence. In such scenarios, $T_2^*$ limits the useful duration of a single experiment, but by exploiting the long-term noise correlations, the noise can be canceled by an echo pulse. Here, we are not interested in applying such echo procedures, since this would also cancel the signal we are trying to measure, but merely point out that successful spin echo protocols prove the existence of long-term noise correlations such that the correlations between shots must be taken into account when determining the overall sensitivity. Long-term noise correlations are usually present in the presence of a slowly evolving nuclear environment or slowly varying external fields and has been reported for e.g. semiconductor quantum dots~\cite{Delbecq2016, struck2020}, superconducting qubits~\cite{Yan2012} and NV centers~\cite{Childress2006, Barbosa2026}. The importance of the noise timescale is further illustrated by a recent NV-ensemble experiment, where the sensitivity gain depended strongly on the dephasing bandwidth~\cite{Gao2026}. Correlations across subsequent experiments are also well established in noise characterization, where they have been used to reconstruct noise spectra~\cite{Fink2013, Malinowski2017, Rojas-Arias2026} and to correct subsequent qubit operations through feedback~\cite{vepsalainen2022, Nakajima2020, Shulman2014}.

Noise correlations across frequency-estimation experiments have been discussed within the quantum-comb framework~\cite{Yang2019, Altherr2021, Das2026}. This framework contains the repeated prepare-evolve-measure protocols discussed here as a restricted case, but also permits coherent memory and adaptive protocols~\cite{Kurdzialek2025_tensor}. Existing applications of quantum combs to sensing with temporally correlated dephasing have focused on noise modeled by simple Markov chains with exponentially decaying correlations~\cite{Kurdzialek2025_tensor, Kurdzialek2025_universal, Ghosh2025}. Rather than treating the number of particles $N$ and the total duration $T$ as separate resources, these works quantify the performance in terms of the number of signal-encoding channel uses $N_{\mathrm{ch}}$. Derived bounds in such scenarios restrict the uncertainty scaling to $1/\sqrt{N_\text{ch}}$, although negative correlations improve the prefactor~\cite{Kurdzialek2025_universal,Ghosh2025}. It therefore remains an open question whether entanglement enables scaling beyond $1/\sqrt{N}$ with the number of probes $N$ under non-Markovian dephasing correlated across repeated experiments.

Beyond temporal correlations, spatial correlations among the $N$ sensing qubits have also been studied extensively. In the case of collective dephasing, where all $N$ probes are subject to the same noise, the estimation uncertainty generally scales unfavorably compared to the spatially uncorrelated noise~\cite{Dorner2012, Beaudoin2018, Riberi2025, Riberi2026}. More general spatial noise correlations have also been investigated in Ramsey spectroscopy. In particular, superclassical scaling has been demonstrated for negatively correlated spatial noise with GHZ and one-axis twisted states in scenarios where the distances between probes are randomized~\cite{Riberi2022} or engineered~\cite{Riberi2023}. 
While Refs.~\cite{Dorner2012, Jeske2014, Beaudoin2018, Riberi2025, Riberi2022, Riberi2023, Riberi2026} discuss various spatial and temporal noise correlations, they all treat successive experimental shots as independent, hence excluding noise correlations that persist across Ramsey sequences. 

In this work, we extend the analysis of entanglement-enhanced quantum sensing in the asymptotic limit of large $N$ and $T$ to noise that is both temporally and spatially correlated, including correlations between experimental shots. We first derive a fundamental limit for quantum sensing, which, in short, states that the signal-to-noise ratio of a sensing experiment cannot exceed the signal-to-noise ratio experienced by the system sensing the signal. This very simple statement is shown to, e.g., capture the sensing limit identified in Ref.~\cite{Huelga1997, Escher2011}. We then demonstrate that, for non-Markovian temporal noise, entanglement can indeed substantially enhance the sensitivity of sensors with entangled probes compared to those with separable probes. In particular, for spin-squeezed states, we show that an uncertainty scaling of $1/N^{3/4}$ is achievable, provided that the environment has a suitable noise spectrum. Finally, we consider combined temporal and spatial noise correlations, showing that anti-correlated spatial noise can support enhanced scaling.

We emphasize that the present study considers quantum sensing, where the goal is to estimate an unknown external field. Although similar techniques, e.g., Ramsey spectroscopy, are often used, this is different from atomic clocks, where the aim is to stabilize a noisy local oscillator. For atomic clocks, the previous discussion therefore does not apply. Instead, it was shown in Refs.~\cite{Kessler2014, Pezze2020, Kielinski2025} that near Heisenberg scaling may be achieved even for Markovian noise.

\section{Fundamental Limit for Frequency Estimation}

We consider a collection of two-level systems with internal levels $\ket{0}$ and $\ket{1}$, and energy splitting $\hbar \sig$ ($\hbar=1$ in the following) induced by an external field $b$ that we want to determine with high precision. A prominent example is spin qubits used to detect magnetic fields via the Zeeman effect. The minimal uncertainty of the obtained estimate $\est$ of the field $\sig$ depends on the number of qubits $N$, the total sensing time $T$, and the noise affecting the system. In this section, we derive a lower bound on the uncertainty in estimating a time-independent frequency $\sig$, assuming stationary, translationally invariant, classical, pure dephasing noise following Gaussian statistics. In particular, we consider the scaling of the uncertainty of a locally unbiased estimator with $N$ particles and over a duration $T$ in the asymptotic limit of large $N$ and $T$.

\subsection{System dynamics}
The system Hamiltonian $H_S$ of an $N$-qubit system with energy splitting $\sig$ is given by 
\begin{equation}
\label{eq:H_system}
    H_S = \frac{\sig}{2} \sum_{n=1}^N \sigma_z^{(n)},
\end{equation}
where $\sigma_k$, with $k = x, y, z$, denote the Pauli matrices. The evolution of the system and its environment is described by the Hamiltonian $H = H_S + H_{SE} + H_E$, where $H_{SE}$ denotes the interaction Hamiltonian with the environment, and $H_E$ the Hamiltonian of the environment. In this work, we restrict to pure dephasing noise, $H_{SE} \propto \sigma_z$, where the system and interaction Hamiltonian commute, $[H_S, H_{SE}] = 0$. This type of 'parallel' noise is considered to be most harmful for frequency estimation~\cite{Haase2018}, and for uncorrelated noise, limits the asymptotic scaling of the estimation uncertainty~\cite{Chaves2013}. 

In the interaction picture with respect to $H_E$, we can represent the interaction Hamiltonian as
\begin{equation}
\Tilde{H}_{SE} = \frac{1}{2} \sum_{n=1}^N \sigma_z^{(n)} \otimes \left[\sig \, \mathbb{I} + F_n(t) \right],
\end{equation}
where $F_n(t)$ are bath operators~\cite{Riberi2022}. Generally, $F_n(t)$ are operators with a non-trivial commutation relation $\left[F_n(0), F_m(t)\right] \neq 0$. In this work, we restrict ourselves to classical environments described by a real, Gaussian random field $\{F_n(t)\}_{n, t}$, and hence $\left[F_n(0), F_m(t)\right] = 0$~\cite{Clerk2010, Szankowski2014}. For a bosonic bath, this assumption is valid in the high-temperature regime, where the occupancy of all relevant noise modes is much larger than unity, which can be expressed as $\beta \omega_c \ll 1$, where $\omega_c$ is the highest relevant noise frequency and $\beta = 1/(k_B T_\text{bath})$ the inverse temperature~\cite{Haase2018, Bing2019}. 
Under the classical pure dephasing assumption, the system dynamics can be described by the effective Hamiltonian $H_{\text{eff}}(t)$~\cite{Bing2019},
\begin{equation}
\label{eq:H_eff}
    H_{\text{eff}}(t) = H_S + \frac{1}{2} \sum_{n=1}^N F_n(t) \sigma_z^{(n)}.
\end{equation}
We assume $F_n(t)$ to be of zero mean, temporally stationary, and spatially translationally invariant, i.e., all probes see statistically equivalent noise and the noise statistics are independent of the absolute time. For simplicity, we consider the probes to be aligned on a line. With these assumptions, the noise correlations $\Ex{F_n(t_1) F_m(t_2)}$ only depend on the distance between them in the emitter chain $(n - m)$ and the time difference $(t_1 - t_2)$~\cite{Jeske2014, Chabuda2020, Riberi2022, Riberi2023}.

\subsection{Fundamental bound}

We consider the system dynamics of Eq.~\eqref{eq:H_eff}. Let the estimator $\est$ be locally unbiased at $\sig$, meaning $\mathbb{E}_{\sig}[\est] = \sig$ and $\partial_{\sig'} \mathbb{E}_{\sig'}\left[{\est}\right]|_{\sig' = \sig} = 1$, with uncertainty $\Delta \est = \sqrt{\Ex{(\est - \sig)^2}}$. In the asymptotic limit $T, N\rightarrow \infty$, with spatial and temporal noise correlation lengths $n_c$ and $t_c$ satisfying $n_c \ll N$ and $t_c \ll T$, this uncertainty is lower bounded by the zero component of the noise spectrum:
\begin{equation}
\label{eq:fund_spatial}
    \Delta \est \geq \sqrt{\frac{D(0, 0)}{N T}},
\end{equation}
where $D(k, \omega)$ is the noise spectrum, defined as the Fourier transform of the noise correlation function
\begin{equation}
\label{eq:noise_spectra}
    D(k, \omega) = \sum_{n = -\infty}^\infty \int_{-\infty}^\infty \eexp{-i (k n + \omega t)} \Ex{F_0(0) F_{n}(t)} dt.
\end{equation}

As a particular example, we first consider uncorrelated spatial and temporal noise. Inserting its autocorrelation function  $\Ex{F_n(t)F_m(t')} = A \, \delta_{nm}\delta(t-t')$, corresponding to a Lindblad operator $c = \sqrt{A}/2 \, \sigma_z$, where $A$ denotes the dephasing strength, into Eq.~\eqref{eq:c00} leads to the bound
\begin{equation}
    \Delta \est \geq \sqrt{\frac{A}{NT}}.
\end{equation}
This corresponds to the bound identified in Refs.~\cite{Huelga1997, Escher2011}.

From Eq.~\eqref{eq:fund_spatial}, we can conclude that for noise spectra with $D(0, 0) \neq 0$, no sensing strategy exceeds a scaling of $1/\sqrt{N}$ in the long-time limit. However, this scaling can always be reached with separable strategies, and hence entanglement does not provide any scaling advantage. For $D(0, 0) = 0$, the bound of Eq.~\eqref{eq:fund_spatial} is trivial and, hence, useless, but it allows scaling beyond $1/\sqrt{N}$.

\subsection{Derivation of the fundamental bound}

The proof of the fundamental bound in Eq.~\eqref{eq:fund_spatial} proceeds by identifying the component of the effective Hamiltonian in Eq.~\eqref{eq:H_eff} that is indistinguishable from the signal $\sig$ and using the variance of this component to lower-bound the variance of $\est$.

Expressing $F_n(t)$ for $t \in [0, T)$ with a Fourier series in time and qubit number yields
\begin{equation}
\label{eq:Heff_FT}
    H_{\text{eff}}(t) \!=\! \sum_{n=1}^N \! \left[\left(\sig \!+\! c_{00}\right) +\!\!\!\!\! \sum_{(s, q) \neq (0, 0)} \!\!\!\! c_{sq} \eexp{2 \pi i \left(\frac{s n}{N} + \frac{q t}{T}\right)} \! \right] \! \frac{\sigma_z^{(n)}}{2},
\end{equation}
where $s \in \{0, 1,..N-1 \}$ and $q \in \mathbb{Z}$. The coefficients $c_{sq}$ are defined by,
\begin{equation}
\label{eq:def_coefficients}
    c_{sq} = \frac{1}{NT} \sum_{n=1}^{N} \int_0^T F_n(t) e^{- 2\pi i \left(\frac{sn}{N} + \frac{qt}{T}\right)}dt.
\end{equation}
As visible in $H_\text{eff}$ of Eq.~\eqref{eq:Heff_FT}, the signal $\sig$ and the noise component $c_{00}$, which is constant in both time and space, affect the system dynamics in exactly the same way. We refer to this constant part of $H_\text{eff}$ as $H_{00}$ and denote by $H_\text{var}$ the part that varies across qubits or time:
\begin{equation}
    H_{00} = \frac{1}{2} \sum_{n=1}^N \left( \sig + c_{00} \right) \sigma_z^{(n)}, \quad H_\text{var} = H_\text{eff} - H_{00}.
\end{equation}

In the asymptotic limit $N,T\rightarrow\infty$, with finite temporal and spatial correlation lengths $t_c$ and $n_c$ satisfying $t_c/T,n_c/N\rightarrow0$, boundary effects become negligible and the covariance $\Ex{F_n(t_1)F_m(t_2)}$ can be treated as periodic~\cite{Gray2006}. The Fourier modes then diagonalize the covariance matrix, implying $\Ex{c_{00} c_{sq}} = 0$ for $(s, q) \neq (0, 0)$~\cite{Gray2006}. Since the Fourier coefficients of a jointly Gaussian random field are themselves jointly Gaussian, their vanishing cross-correlations imply that $c_{00}$ is statistically independent of all other Fourier coefficients. Consequently, $H_\text{var}$ contains no information about $H_{00}$ and, by the data-processing inequality, cannot improve the estimation precision~\cite{Nielsen2000}. If it would be an advantage to have the extra noise term $H_\text{var}$, it can always be included as a part of the control strategy by adding noise to the system. We can therefore lower-bound the estimation uncertainty by considering $H_{00}$ alone.

We now regard $x = \sig + c_{00}$ as the effective parameter of $H_\text{00}$. The information processing inequality states that any parameter-independent measurement on $x$ cannot contain more information about $\sig$ than $x$ itself~\cite{Nielsen2000}. The optimal measurement would therefore extract $x$ directly. For zero-mean Gaussian-distributed $c_{00}$ and a locally unbiased estimator $\est$ of $\sig$ based on $x$, this yields,
\begin{equation}
\label{eq:fund_lim_omega}
    \Var{\est} \geq \Var{x} = \Var{c_{00}} = \Ex{c_{00}^2}.
\end{equation}

The variance of the zero-frequency noise component $\Ex{c_{00}^2}$ is, according to Eq.~\eqref{eq:def_coefficients},
\begin{equation}
\label{eq:c00}
    \Var{c_{00}} \!=\! \frac{1}{(N T)^2} \sum_{n=1}^{N} \sum_{m=1}^{N} \int_0^T \!\!\! \! \int_0^T \!\!\!\! \! \Ex{F_{n}(t) F_{m}(t')} dt dt'.
\end{equation}
In the asymptotic limit of $N, T \rightarrow \infty$ and stationarity and translational invariance give,
\begin{equation}
    \Var{c_{00}} = \frac{D(0, 0)}{NT}.
\end{equation}
Using this expression, Eq.~\eqref{eq:fund_lim_omega} leads directly to the fundamental bound in Eq.~\eqref{eq:fund_spatial}. 

~\\

\section{Noisy Ramsey spectroscopy with spin-squeezed states}
\label{sec:uncertainty_derivation}

\begin{figure*}[]
  \centering
  \includegraphics[width=0.85\textwidth]{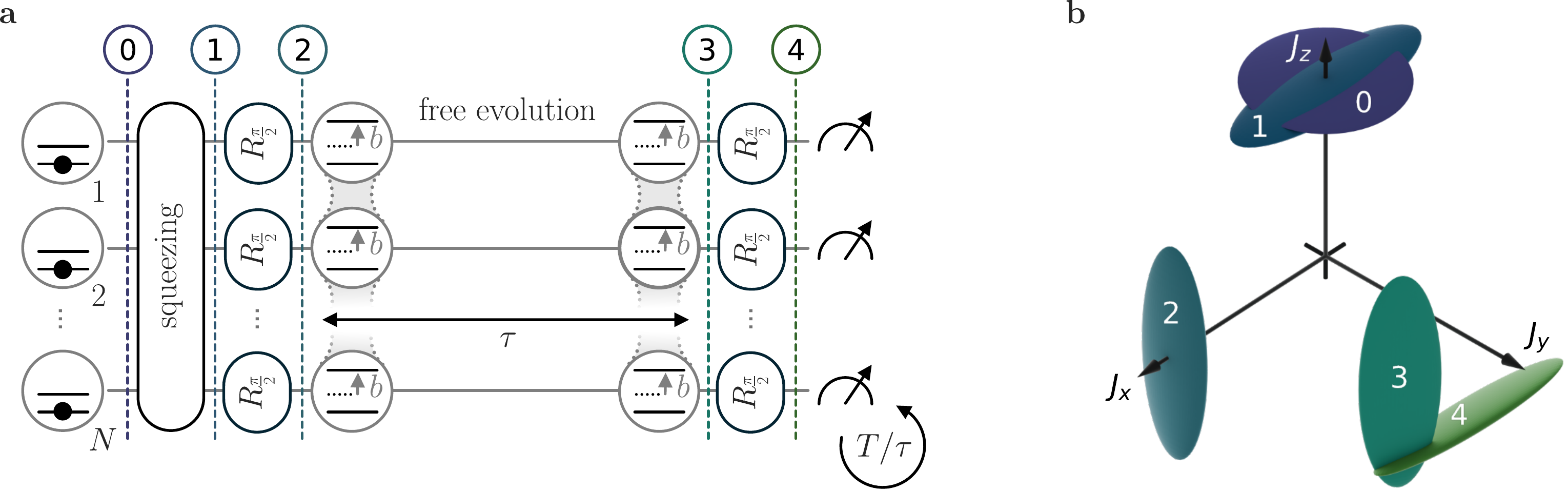}
  \caption{(a) Circuit representation of Ramsey spectroscopy using spin-squeezed states. Applying squeezing to the $N$-qubit ground state reduces $\Ex{J_y^2}$ and, to a lesser degree, $\Ex{J_z}$. Next, a $\pi/2$ pulse ($R_{\pi/2}$) rotating the collective spin around the $J_y$-axis is applied. The state subsequently freely evolves for a time $\tau$, during which it accumulates a relative phase proportional to the energy splitting. This phase is finally estimated by applying another $\pi/2$ pulse and a measurement in the $J_z$-basis. The total procedure is repeated $T/\tau$ times. (b) Schematic representation of the quantum state at different stages of the protocol, shown on the Bloch sphere. The ellipsoids are centered at the expectation values of the spin operators $\Ex{J_k}$, with $k = x, y, z$, and the axis length represents their uncertainty $\Delta J_k$. The disc labeled with $0$ depicts the unsqueezed state with $\Delta J_z = 0$ and $\Delta J_x = \Delta J_y$.  The Ramsey pulses correspond to a rotation of the ellipsoids by an angle of $\pi / 2$ around the $J_y$ axis. The phase accumulation during the free evolution results in a rotation around the $J_z$ axis. The uncertainty of the final measurement in the $J_z$ basis is given by the width of ellipsoid $4$ along the $J_z$ direction. This illustrates how squeezing the initial state reduces shot noise and thereby improves the precision of the phase estimation.
}
  \label{fig:fig1}
\end{figure*}

Having established a fundamental limit, we now turn to analyzing the performance of concrete protocols. There exists a wide variety of protocols for measuring a static energy splitting in an $N$-qubit system within a fixed total interrogation time $T$, particularly when entangling operations are permitted during state preparation and measurements~\cite{Giovannetti2006}. Quantum information theoretic tools, such as quantum Fisher information, have been used to derive general bounds on the sensitivity for situations where different shots of the experiment are uncorrelated~\cite{Escher2011, Demkowicz2012, Matsuzaki2011, Chin2012, Macieszczak2015, Smirne2016, Haase2018}. Incorporating correlations across experiments into the Fisher information is challenging due to the exponential growth in the number of possible measurement outcome sequences~\cite{Yang2023, OConnor2024}. Methods capable of handling this complexity have been developed in the context of sequential measurement metrology and studied, for example, in quantum thermometry~\cite{Seah2019, Yang2023, Yang2024, OConnor2024}. Within the quantum comb framework, the quantum Fisher information has been numerically evaluated using tensor networks~\cite{Kurdzialek2025_tensor}, bounded generally~\cite{Kurdzialek2025_universal}, and studied for correlated dephasing~\cite{Ghosh2025}. Here, we restrict ourselves to a specific protocol with a limited set of parameters to quantify the improvement in estimation uncertainty provided by entanglement under classical pure dephasing noise. We follow the protocol of Ref.~\cite{Wineland1994}, which corresponds to standard Ramsey spectroscopy~\cite{Degen2017}, but with the qubits being initialized in a spin-squeezed state rather than the separable state $\ket{0}^{\otimes N}$. Furthermore, we consider the local estimation regime, where the frequency $\sig$ is approximately known~\cite{Haase2018}, which is the most common paradigm. To quantify the entanglement enhancement, we discuss the metrological gain
\begin{equation}
\label{eq:r}
    r = \frac{\Delta \est \vert_{\text{sep}}}{\Delta \est \vert_{\text{ent}}}   
\end{equation} 
comparing the uncertainty of standard Ramsey spectroscopy (separable), $\Delta \est \vert_{\text{sep}}$, with that of the entangled protocol, $\Delta \est \vert_{\text{ent}}$. We consider the estimators to be unbiased $\Ex{\est} = b$, which implies that the estimator's variance corresponds to its mean square error $\Var{\est} = \Ex{(\est - b)^2}$.

The advantage of using spin-squeezed states becomes evident from the uncertainty expression for noiseless conventional Ramsey spectroscopy. The uncertainty depends on the initial variance $\Delta J_y$  and the initial expectation value $\Ex{J_z}$ according to $\Delta \est \propto \Delta J_y / \Ex{J_z}$~\cite{Wineland1992}, where $J_k = 1/2 \Sigma_n \sigma_k^{(n)}$, with $k = x, y, z$, is the spin operator in the $k$-direction. For separable input states, $\Delta J_y / \Ex{J_z} \geq 1 / \sqrt{N}$ leads to a frequency estimation uncertainty scaling as $1/\sqrt{N}$, known as the shot-noise limit, since it arises from the probabilistic nature of spin measurements. Spin-squeezed states~\cite{Kitagawa1993, Wineland1992, Wineland1994} allow $\Delta J_y / \Ex{J_z} < 1/\sqrt{N}$ and can, assuming no other sources of noise, exceed the shot-noise limit. Moreover, spin-squeezed states are highly symmetric states with controllable entanglement. This is advantageous since, for Markovian dephasing, it was shown that symmetric states with limited entanglement outperform the more entangled GHZ states~\cite{Huelga1997}. For these reasons, spin-squeezed states are a well-established resource for demonstrating quantum-enhanced Ramsey spectroscopy~\cite{Wineland1994, Huelga1997, Borregaard2013, Riberi2023}, analogous to the use of photonic squeezed states in Mach-Zehnder interferometry~\cite{Giovannetti2004}.

The protocol considered, illustrated in Fig.~\ref{fig:fig1}, consists of the following steps:
\begin{enumerate}[noitemsep]
    \item Preparation of a spin-squeezed state;
    \item $\pi/2$ pulse around the collective spin $J_y$-axis;
    \item Free evolution of the system for a time $\tau$;
    \item $\pi/2$ pulse around the collective spin $J_y$-axis;
    \item Measurement of the collective spin component $J_z$;
    \item Repetition of steps 1-5 for $L = T/\tau$ repetitions.
\end{enumerate}
This description is in the rotating frame of the field that implements the $\pi/2$ Ramsey pulses, which rotates around the $J_z$-axis with tunable frequency $\sig_p$. In this frame, the relative phase between the excited and ground states evolves at a frequency $\sig_r = \sig - \sig_p$. We assume that the duration of state preparation, measurement, and pulse operations is negligible compared to the free evolution time $\tau$, such that the total interrogation time is effectively $T = L \tau$.

Throughout the $L$ experiments, we consider that the state preparation and the evolution time $\tau$ are kept constant. This leaves three free experimental parameters: the amount of squeezing, the evolution time $\tau$, and the operating point of the sensor $\sig_r \tau$. To parametrize the squeezing with a single parameter, $\kappa$, we consider the following class of states~\cite{Borregaard2013}:
\begin{equation}
\label{eq:squeezed_states}
    \ket{\psi_\kappa} = \mathcal{N}(\kappa) \sum_{m=-N/2}^{N/2} (-1)^m \eexp{- \frac{m^2}{N \kappa^2}} \ket{m}_y,
\end{equation}
where $\kappa \in (0, 1]$ is the squeezing parameter, $\mathcal{N}(\kappa)$ the normalization and $\ket{m}_y$ the Dicke state $\ket{N/2, m}_y$ in the $y$-basis. For $\kappa = 1$, up to an error in approximating the binomial distribution with a Gaussian, we recover a coherent spin state. For $\kappa < 1$ we have a spin-squeezed state with $\Delta J_y/\Ex{J_z} < 1/\sqrt{N}$. From the state in Eq.~\eqref{eq:squeezed_states}, any expectation values of the spin operators can be calculated. The states are symmetric with respect to particle exchange and $\Ex{J_y} = \Ex{J_x} = 0$. We consider these states since they form a simple class that allows easy evaluation of all relevant expectation values. The states produced experimentally may, however, not have this form. In App.~\ref{app:spinsqueezing} we compare these states to those produced by the commonly considered one- and two-axis squeezing. These are found to perform similarly, except in certain situations, where the one-axis twisting may not be able to produce sufficient amounts of squeezing.

For the derivation of the uncertainty of the estimation, we operate in the Heisenberg picture and use the effective Hamiltonian of Eq.~\eqref{eq:H_eff}. The measurement operator $M_l(\tau)$ of the $l$-th experiment is derived by evolving the spin operators through the Ramsey sequence to the time of measurement, 
\begin{equation}
\begin{aligned}
    M_{l}(\tau) = \sum_{n = 1}^N & \Bigl[-\cos\left({\sig_r \tau + \phi_{nl\tau}}\right) S_{z}^{(ln)}(0) \\
    & + \sin\left({\sig_r \tau + \phi_{nl\tau}}\right) S_{y}^{(ln)}(0) \Bigr],
\end{aligned}
\end{equation}
where $\phi_{nl\tau} = \int_{l\tau}^{(l+1)\tau} \! F_n(t) \,dt$ and $S_k^{(ln)}(0)$ are the spin operators ($S_k = \sigma_k/2$) of qubit $n$ before the first Ramsey pulse of experimental shot $l$. We assume that $\phi_{nl\tau}$ has a Gaussian distribution with zero mean. Unlike previous treatments of colored noise~\cite{Matsuzaki2011, Chin2012}, we consider the noise to be correlated between different experimental shots and qubits, meaning $\Ex{\phi_{nl\tau}\phi_{n'l'\tau}}$ is not necessarily $0$ for $l \neq l'$ or $n \neq n'$. 

The response of the measurement outcome to changes in the signal $\abs{\partial \Ex{M_l(\tau)}/\partial\sig}$ is maximal for $\sig_r \tau = q \, \pi/2$, with odd $q$. We consider operating the sensor around this point of optimal response and denote the deviation from the optimal point by the phase $\theta$,
\begin{equation}
\begin{aligned}
\label{eq:measurement_theta}
    M_{l}(\tau) = \sum_{n = 1}^N & \Bigl[ \sin\left({\theta + \phi_{nl\tau}}\right) S_{z}^{(ln)}(0) \\
    & + \cos\left({\theta + \phi_{nl\tau}}\right) S_{y}^{(ln)}(0) \Bigr].
\end{aligned}
\end{equation}
Operating in the local estimation regime, meaning  $\theta \ll 1$, we linearize the relation between $\sig$ and $\Ex{M_l(\tau)}$. With these assumptions, the frequency estimate $\est$ is obtained from the measurement outcomes ${m_l}$ according to 
\begin{equation}
\label{eq:estimation}
    \est = \frac{1}{L} \sum_{l=0}^{L-1} \frac{m_l}{\frac{\partial \Ex{M_l(\tau)}}{\partial \sig}} + \sig_p + \frac{q \pi}{2 \tau},
\end{equation}
where $\sig_p$ is the frequency of the rotating frame. To ensure that the estimator is unbiased~\cite{Riberi2023}, we consider the case where the noise-dependent slope ${\partial \Ex{M_l(\tau)}}/{\partial \sig}$ is known from sensor calibration and assume that the experiments are affected by the same noise statistics as during calibration. 

The estimation uncertainty $\Delta \est$ is given by the ratio of the standard deviation of the total measurement $M(\tau) = \sum_l M_l(\tau)$ and the response of $\Ex{M(\tau)}$ to changes in $\sig$~\cite{Pezze2018},
\begin{equation}
\label{eq:uncertainty_general}
    \Delta \est = \frac{\Delta M(\tau)}{L \abs{\frac{\partial \Ex{M_0(\tau)}}{\partial \sig}}},
\end{equation} 
where we used that for stationary noise $\Ex{M_l(\tau)}$ is independent of $l$. To evaluate Eq.~\eqref{eq:uncertainty_general} with Eq.~\eqref{eq:measurement_theta} we do a first-order approximation in $\theta$,
\begin{equation}
\begin{aligned}
\sin\left({\theta\!+\!\phi_{nl\tau}}\right) &\approx \frac{(1\!+\!i \theta) \eexp{i \phi_{nl\tau}} \!-\! (1\!-\!i \theta) \eexp{-i \phi_{nl\tau}}}{2i}\\
\cos\left({\theta\!+\!\phi_{nl\tau}}\right) &\approx \frac{(1\!+\!i \theta) \eexp{i \phi_{nl\tau}} \!+\! (1\!-\!i \theta) \eexp{-i \phi_{nl\tau}}}{2}.
\end{aligned}
\end{equation}

Using that, for a Gaussian random variable $X$, $\Ex{\eexp{i X}} = \eexp{-\Ex{X^2}/2}$, and keeping terms up to first order in~$\theta$, we obtain
\begin{equation}
\begin{aligned}
\Ex{\sin\!\left({\theta\!+\!\phi_{00\tau}}\right) \!\sin\!\left({\theta\!+\!\phi_{nl\tau}}\right)} & \! \approx \! \eexp{-\!\gamma(\tau)} \!\sinh{\!\Ex{\phi_{00\tau} \phi_{nl\tau}}}\\
\Ex{\cos\!\left({\theta\!+\!\phi_{00\tau}}\right) \!\cos\!\left({\theta\!+\!\phi_{nl\tau}}\right)} & \! \approx \! \eexp{-\!\gamma(\tau)} \!\cosh{\!\Ex{\phi_{00\tau} \phi_{nl\tau}}},
\end{aligned}
\end{equation}

\noindent where the dephasing is contained in the function $\gamma(\tau) = \Ex{\phi_{nl\tau}^2}$. These equations, combined with the properties of the introduced spin-squeezed states $\Ex{J_y(0)} = \Ex{J_x(0)} = 0$ and the assumption  that the spins are initialized at each shot such that the spins of different experiments are uncorrelated, lead to the estimation uncertainty
\begin{widetext}
\begin{equation}
\label{eq:uncertainty_total}
    \begin{aligned}
         \Var{\est} = \biggl\{ & \frac{N L}{4} \eexp{\gamma(\tau)} - \frac{L}{4 (N-1)} \sum_{n \neq n'}^{N} \eexp{\Ex{\phi_{n0\tau} \phi_{n'0\tau}}} + \frac{\Ex{J_z(0)}^2}{N^2} \sum_{n, n'}^{N} \sum_{l \neq l'}^L \sinh\left(\Ex{\phi_{nl\tau} \phi_{n'l'\tau}} \right)\\
         &+ \frac{L}{N (N-1)} \sum_{n \neq n'}^{N} \left[ \Ex{{J_z(0)}^2} \sinh\left(\Ex{\phi_{n0\tau} \phi_{n'0\tau}}\right) + \Ex{{J_y(0)}^2} \cosh\left( \Ex{\phi_{n0\tau} \phi_{n'0\tau}} \right) \right] \biggr\} \biggl\{ T \Ex{J_z(0)} \biggr\}^{-2}.
    \end{aligned}  
\end{equation}
\end{widetext}
This expression will form the basis of our analysis of the estimation uncertainty.

\section{Uncertainty Analysis}
\label{sec:uncertainty_analysis}
 
In this section, we discuss Eq.~\eqref{eq:uncertainty_total} for different spatial and temporal noise spectra. We start by briefly discussing the well-established scenarios of no- and uncorrelated noise. We then turn to non-Markovian temporal and spatially correlated noise with correlations between different shots of the experiment.

\subsection{No noise}
By setting $\phi_{nl\tau} = 0$ in Eq.~\eqref{eq:uncertainty_total} we recover the well-known estimation uncertainty in the absence of noise~\cite{Wineland1992},
\begin{equation}
\label{eq:uncertainty_nonoise}
   \Delta \est = \frac{\Delta J_y(0)}{\sqrt{T \tau} \Ex{J_z(0)}}.
\end{equation}
The presence of uncertainty in estimating $\sig$, even in the absence of external noise, arises from the probabilistic nature of measurements. For separable input states, the ratio $\Delta J_y/\Ex{J_z}$ is lower bounded by $1/\sqrt{N}$, leading to a scaling of $\Delta \est = 1/\sqrt{\tau T N}$, known as the shot noise limit~\cite{Itano1992}, while for spin-squeezed states this ratio is lowered and can reach the lower bound $1/N$~\cite{Ma2011}. Consequently, spin squeezing improves the uncertainty scaling in the absence of noise~\cite{Wineland1992}.

\subsection{Uncorrelated noise}
\label{subsec:markoviannoise}
Temporally uncorrelated noise, called white or Markovian, which is also spatially uncorrelated, is characterized by $\Ex{\phi_{nk\tau} \phi_{n'k'\tau}} = A \tau \, \delta_{nn'} \, \delta_{kk'}$ and consequently $\gamma(\tau) = A \tau$. Under these noise properties, Eq.~\eqref{eq:uncertainty_total} simplifies to
\begin{equation}
\label{eq:uncertainty_markovian}
    \Var{\est} = \frac{\frac{N}{4}\left(\eexp{A \tau} - 1 \right) + \Ex{J_y(0)^2}}{T \tau \Ex{J_z(0)}^2}.
\end{equation}

\begin{figure}[]
  \centering
  \includegraphics[width=\columnwidth]{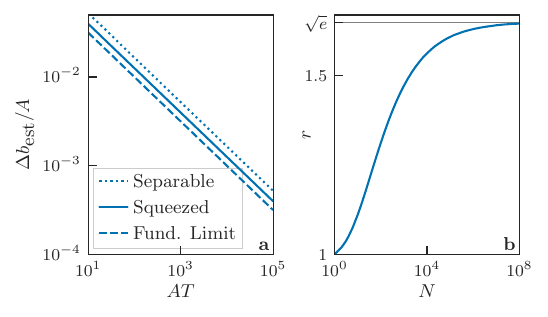}
\caption{Comparison of estimation uncertainty in Ramsey spectroscopy using separable and spin-squeezed input states in the presence of spatially and temporally uncorrelated noise characterized by a strength $A$. (a) The estimation uncertainty $\Delta \est / A$ as a function of the total time $AT$ for a fixed number of qubits $N = 100$. The dotted line indicates unsqueezed states, the full line corresponds to squeezed states, and the dashed line indicates the fundamental limit. All curves scale with $1/\sqrt{T}$ and differ only by constant factors. (b) The metrological gain $r$ between the uncertainties of the spin-squeezed and separable protocols at fixed $AT = 1000$ as a function of $N$. As $N$ increases, the uncertainty of the spin-squeezed protocol approaches the fundamental limit of Eq.~\eqref{eq:fund_spatial}, yielding an asymptotic improvement by a factor of $\sqrt{e}$ compared to the separable protocol.}
  \label{fig:fig2}
\end{figure}

For separable input states, with $\Ex{J_z(0)} = N/2$ and $\Ex{J_y(0)^2} = N/4$, the optimal interrogation time is given by $\tau = 1/A$, leading to an estimation uncertainty of $\Delta\est \vert_\text{sep} = \sqrt{Ae}/\sqrt{NT}$. This sets the minimal achievable uncertainty for all separable strategies~\cite{Huelga1997}. By introducing spin-squeezed input states, the optimization involves both the squeezing parameter $\kappa$ and the interrogation time $\tau$. While entanglement does not alter the $1/\sqrt{T}$ scaling, increasing $N$ enhances the metrological gain $r$ defined in Eq.~\eqref{eq:r}, as illustrated in Fig.~\ref{fig:fig2}. However, $r$ cannot exceed $\sqrt{e}$ due to the fundamental limit of Eq.~\eqref{eq:fund_spatial}, and hence, in the case of Markovian dephasing noise, entanglement only offers a constant-factor enhancement over separable strategies~\cite{Escher2011}. Notably, $r=\sqrt{e}$ cannot be reached with maximally entangled GHZ states, but it can be approached using spin-squeezed states, which are highly symmetric yet only partially entangled~\cite{Huelga1997, Orgikh2001}.  

\subsection{Temporally correlated and spatially uncorrelated noise}
\label{subsec:tempcor_spatmark}

In this subsection, we discuss general temporally correlated but spatially uncorrelated noise, i.e., 
\begin{equation}
\label{eq:spatial_uncor_noise}
    \Ex{\phi_{nl\tau}\phi_{n'l'\tau}} = \delta_{nn'} \,  Q_\tau(l-l').
\end{equation}
Given a temporal noise spectrum $S(\omega)$ we can evaluate $Q_\tau(\Delta_l)$ to be~\cite{Uhrig2008, Biercuk2011}
\begin{equation}
\label{eq:formula_noise_correlations}
    Q_\tau(\Delta_l) \!=\! \frac{\tau^2}{\pi} \! \int_{0}^{\infty} \!\!\!\! S(\omega) \, \text{sinc}\!\left(\frac{\tau \omega}{2} \right)^2 \!\! \cos{\!\left( \tau \omega \Delta_l \right)} \, d\omega,    
\end{equation}
where the $\text{sinc}$-function emerges from the Fourier transform of the rectangular time window of a single Ramsey sequence and $\Delta_l = l-l'$. 

\begin{figure*}[ht]
  \centering
  \includegraphics[width=\textwidth]{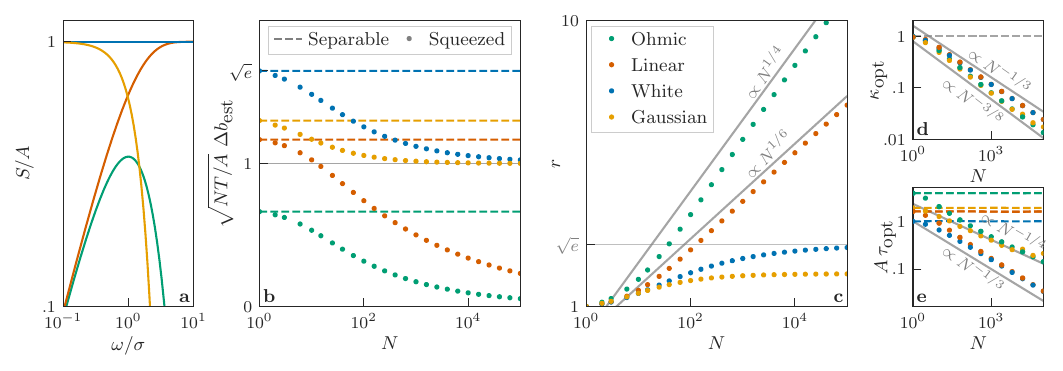}
  \caption{Estimation uncertainty of Ramsey spectroscopy with separable and spin-squeezed input states for different noise spectra and a certain total probing time $T$ much larger than the noise correlation time and dephasing time. (a) The considered noise spectra. We choose $\sigma / A = 1/2$. (b) The uncertainty scaled with $\sqrt{NT/A}$ as a function of the number of qubits $N$ for spin-squeezed and separable input states under noise with (counting from below at $N=1$) Ohmic, linear, Gaussian, and white spectra. For the Gaussian and white noise, the uncertainty approaches the fundamental limit of Eq.~\eqref{eq:fund_spatial}, which is $ \Delta \est = \sqrt{A/NT}$. For the linear and Ohmic noise, on the other hand, the noise is zero at zero frequency, meaning that Eq.~\eqref{eq:fund_spatial} does not pose a constraint. We therefore find a large improvement from using spin-squeezed states. (c) The metrological gain $r$, corresponding to the ratio of the estimation uncertainty with and without squeezing. The optimal squeezing parameter $\kappa$ and interrogation time $\tau$ are displayed in (d) and (e).}
  \label{fig:fig3}
\end{figure*}

In addition to the Markovian noise spectrum $S_W(\omega)$, we consider three non-Markovian noise spectra, each characterized by a noise strength $A$ and a characteristic frequency $\sigma$:
\begin{equation}
\begin{aligned}
\label{eq:spectrum_tempnoise}
    S_W(\omega) &= A, \\ S_G(\omega) &= A \eexp{-\frac{\omega^2}{2\sigma^2}}, \\
    S_{L}(\omega) &= A \left(1 - \eexp{-\frac{\abs{\omega}}{\sigma}}\right), \\S_{O}(\omega) &= A \frac{\abs{\omega}}{\sigma} e^{-\frac{\abs{\omega}}{\sigma}}.
\end{aligned}
\end{equation}
The noise with a Gaussian spectrum $S_G(\omega)$ represents correlated noise with a correlation time $1/\sigma$, as its autocorrelation function is also Gaussian. The linear spectrum $S_L(\omega)$, which is linear at small frequencies and flat at high frequencies, is an example of temporally anti-correlated noise. Its autocorrelation function consists of a delta peak at $t=0$ and a negative Lorentzian with a time scale $1/\sigma$. Finally, the Ohmic spectrum $S_O(\omega)$ corresponds to noise that is correlated at short times and anti-correlated at longer times. An example of the spectra is illustrated in Fig.~\hyperref[fig:fig3]{\ref{fig:fig3}a}. The autocorrelation function and the dephasing $\Ex{\phi_{0 0 \tau}^2} = \gamma(\tau)$ for the noise spectra of Eq.~\eqref{eq:spectrum_tempnoise} are given in App.~\ref{app:dephasingrate}. We note that the terms "Gaussian", "linear", and "Ohmic" in Eq.~\eqref{eq:spectrum_tempnoise} refer only to the frequency dependence of the power spectra. This should not be confused with Gaussian statistics, which is assumed throughout this paper.

Simplifying Eq.~\eqref{eq:uncertainty_total} for spatially uncorrelated noise leads to
\begin{equation}
\label{eq:uncertainty_temp_unc}
\begin{aligned}
   \Var{\est} =& \frac{\Ex{{J_y(0)}^2}}{\tau T \Ex{J_z(0)}^2} + \frac{N \left[\eexp{\gamma(\tau)} - 1\right]}{4 \tau T \Ex{J_z(0)}^2} \\
   &+ \frac{\sum_{l \neq l'}^L \sinh{\left[\Ex{\phi_{0 l \tau}\phi_{0 l' \tau}}\right]}}{N T^2}.
\end{aligned}
\end{equation}
The last term in Eq.~\eqref{eq:uncertainty_temp_unc} accounts for correlations among experimental shots. Leaving this term out corresponds to the scenario in which the noise is reset after each Ramsey sequence, as discussed in Refs.~\cite{Matsuzaki2011, Chin2012, Macieszczak2015}.

For a discussion of Eq.~\eqref{eq:uncertainty_temp_unc}, but not for the displayed numerical results, we expand $\sinh x = x+O(x^3)$ and drop the higher order terms, assuming $\Ex{\phi_{0k\tau}\phi_{0k'\tau}}$ to be small, which leads to
\begin{equation}
\label{eq:uncertainty_temp_simp}
   \Var{\est\!} \!\! \approx \!\!\frac{\Ex{\!{J_y(0)}^2}\!}{\tau T \!\Ex{J_z(0)}^{\!2}} + \frac{N \! \left[\eexp{\gamma(\tau)} \!-\! 1\right]}{4 \tau T \! \Ex{J_z(0)}^{\!2}} - \frac{\gamma(\tau)}{N \tau T} + \frac{\gamma(T)}{N T^2}.
\end{equation}

The last term of Eq.~\eqref{eq:uncertainty_temp_simp} is independent of the precise quantum state of the system and the interrogation time $\tau$. In the long-time limit $T \rightarrow \infty$, $\Delta\est$ scales with $1/\sqrt{T}$ and $\gamma(T)/T$ can be approximated by the zero-frequency noise component $S(0)$ (see App.~\ref{app:timediscussion}). As the sum of the first three terms in Eq.~\eqref{eq:uncertainty_temp_simp} is positive, we immediately recover the fundamental bound of Eq.~\eqref{eq:fund_spatial} from the last term. The $N$-scalings discussed below apply in the long-time limit, meaning that for all considered $N$ we assume $T$ to be sufficiently large for finite-time contributions, such as $\gamma(T)/(NT^2)$ when $S(0)=0$, to remain negligible.

Unlike the last term, the sum of the first three terms on the right-hand side of Eq.~\eqref{eq:uncertainty_temp_simp} depends on the squeezing parameter $\kappa$ and the interrogation time $\tau$. We consider the long-time limit such that the optimal duration of a single experiment $\tau_{\text{opt}}$ satisfies $\tau_{\text{opt}} \ll T$, i.e., where we perform many Ramsey sequences. Decreasing $\kappa$, which corresponds to introducing more entanglement, reduces the first term associated with the shot noise as described in Eq.~\eqref{eq:uncertainty_nonoise}. However, additional squeezing also suppresses $\Ex{J_z(0)}$ and thus increases the second term. The interrogation time $\tau$ introduces a different tradeoff. Reducing $\tau$ increases the shot-noise term but decreases the sum of the second and third terms, which are linked to dephasing. The interplay between these effects determines the ideal parameters. The optimal values of $\kappa$ and $\tau$ depend mainly on the short-time behavior of $\gamma(\tau)$, which determines how the second and third terms scale with $\tau$. Finally, a larger number of qubits $N$ enables stronger squeezing and a shorter optimal interrogation time, leading to a lower estimation uncertainty.

In Fig.~\hyperref[fig:fig3]{\ref{fig:fig3}b}, we show the minimal uncertainty of Eq.~\eqref{eq:uncertainty_temp_unc}, obtained by numerically optimizing both the squeezing and the interrogation time for the four noise spectra defined by Eq.~\eqref{eq:spectrum_tempnoise}. We rescale the obtained result such that for separable states, we get flat lines. Introducing entanglement through squeezing improves the uncertainty for higher $N$ compared to the separable case. As visible in Fig.~\hyperref[fig:fig3]{\ref{fig:fig3}e}, the optimal interrogation time decreases with increasing $N$. Simultaneously, the squeezing parameter $\kappa$ also decreases, as shown in Fig.~\hyperref[fig:fig3]{\ref{fig:fig3}d}, corresponding to stronger entanglement. For the noise spectra with a nonzero zero-frequency noise component, i.e., Gaussian and white spectra, Eq.~\eqref{eq:fund_spatial} limits the normalized uncertainty to be greater than $1$ in the rescaled parameters. In contrast, for the linear and Ohmic spectra, no such limit exists, allowing for a substantial improvement of the estimation uncertainty. This behavior is also visible in Fig.~\hyperref[fig:fig3]{\ref{fig:fig3}c}, where the gain $r$ saturates with increasing $N$ for white and Gaussian noise, while it remains $N$ dependent for Ohmic and linear noise.

To quantify the observed entanglement-enhancement, we consider a generic time dependence $\gamma(\tau) \propto \tau^\nu$. Using the approximate uncertainty of Eq.~\eqref{eq:uncertainty_temp_simp} we can find the optimal $\tau$ and $\kappa$ by solving $\partial_\tau \Delta \est = 0$ and $\partial_\kappa \Delta \est = 0$. In leading order, we find $\tau_{\text{opt}} \propto N^{-1/3}$ and $\kappa_{\text{opt}} \propto N^{-1/3}$ for $\nu = 1$, and $\tau_{\text{opt}} \propto N^{-1/4}$ and $\kappa_{\text{opt}} \propto N^{-3/8}$ for $\nu = 2$. For spectra with $S(0)=0$ this leads to an uncertainty scaling as $\Delta \est \propto N^{-2/3}$ for $\nu = 1$ ($r \propto N^{1/6}$), and  $\Delta \est \propto N^{-3/4}$ for $\nu = 2$ ($r \propto N^{1/4}$). These findings, summarized in Tab.~\ref{tab:noise-scaling}, agree with the observed scaling of $r$, $\tau_{\text{opt}}$, and $\kappa_{\text{opt}}$ in Fig.~\hyperref[fig:fig3]{\ref{fig:fig3}c-e}. There we see that among the noise spectra where the uncertainty is not bounded by the fundamental limit of Eq.~\eqref{eq:fund_spatial}, a better scaling is achievable for Ohmic noise, $r \propto N^{1/4}$, than for linear noise, $r \propto N^{1/6}$. This scaling difference can be attributed to the different short-time noise behavior, for the Ohmic spectrum $\gamma(\tau) \propto \tau^2$, while for the linear spectrum $\gamma(\tau) \propto \tau$. Hence, lowering the interrogation time in the case of Ohmic noise decreases the dephasing within a single shot more than for the linear noise. Through the above-described tradeoff between $\kappa$ and $\gamma(\tau)$, this leads to a better uncertainty scaling for Ohmic noise spectra.

\begin{table}[htbp]
\renewcommand{\arraystretch}{1.4}
\begin{tabular}{l||c|c|c|c|c}
& $\gamma(\tau)$ & $\gamma(T)$ & $\Delta \est$ & $\tau_{\text{opt}}$ & $\kappa_{\text{opt}}$ \\
\hline
\hline
White & $\propto \tau$ & $\propto T$ & $\propto N^{-1/2}$ & $\propto N^{-1/3}$ & $\propto N^{-1/3}$ \\
\hline
Gaussian & $\propto \tau^2$ & $\propto T$ & $\propto N^{-1/2}$ & $\propto N^{-1/4}$ & $\propto N^{-3/8}$ \\
\hline
Linear & $\propto \tau$ & $\propto \log(T)$ & $\propto N^{-2/3}$ & $\propto N^{-1/3}$ & $\propto N^{-1/3}$ \\
\hline
Ohmic & $\propto \tau^2$ & $\propto \log(T)$ & $\propto N^{-3/4}$ & $\propto N^{-1/4}$ & $\propto N^{-3/8}$ \\
\end{tabular}
\caption{Leading order scaling of short-time $\gamma(\tau)$ and long-time dephasing, $\gamma(T)$, estimation uncertainty $\Delta \est$, optimal interrogation time $\tau_\text{opt}$, and optimal squeezing $\kappa_\text{opt}$ with time, respectively number of qubits $N$ across different noise spectra.}
\label{tab:noise-scaling}
\end{table}

To summarize this analysis, the noise spectrum affects the estimation uncertainty in two ways. First, the zero-frequency component of the spectrum sets a limit on the achievable uncertainty and thus determines whether an uncertainty scaling of $1/\sqrt{N}$ can be surpassed. Second, for noise spectra with $S(0) = 0$, the short-time behavior of $\gamma(\tau)$, which reflects the high-frequency characteristics of the spectrum, governs the extent to which spin squeezing can enhance the scaling of the estimation uncertainty with entangled states. Overall, entanglement improves the measurement uncertainty in all the cases considered. In the large-$N$ limit, the smallest improvement occurs for correlated noise with $S(0) \neq 0$, while the largest improvement occurs for noise with $S(0) = 0$, with $\gamma(\tau) \propto \tau^2$.

Compared to the noise model and protocol proposed in Refs.~\cite{Matsuzaki2011, Chin2012}, where the noise among different Ramsey sequences is considered to be uncorrelated, we observe two key differences. First, the scaling advantage reported for the Gaussian spectra~\cite{Chin2012} is absent in our case, and likewise would be absent for a Lorentzian spectrum~\cite{Matsuzaki2011}, because of the non-zero zero-frequency component of the noise spectrum. Second, unlike in Refs.~\cite{Matsuzaki2011, Chin2012}, our reported scaling advantage does not depend on the superlinear scaling of $\gamma(\tau)$ in $\tau$ because we use spin-squeezed states instead of GHZ states (see App.~\ref{app:GHZ}). However, we also report the biggest scaling advantage in the cases when the noise is correlated at short times and hence $\gamma(\tau) \propto \tau^2$.

\subsection{Temporally and spatially correlated noise}
\begin{figure*}[ht]
  \centering
  \includegraphics[width=\textwidth]{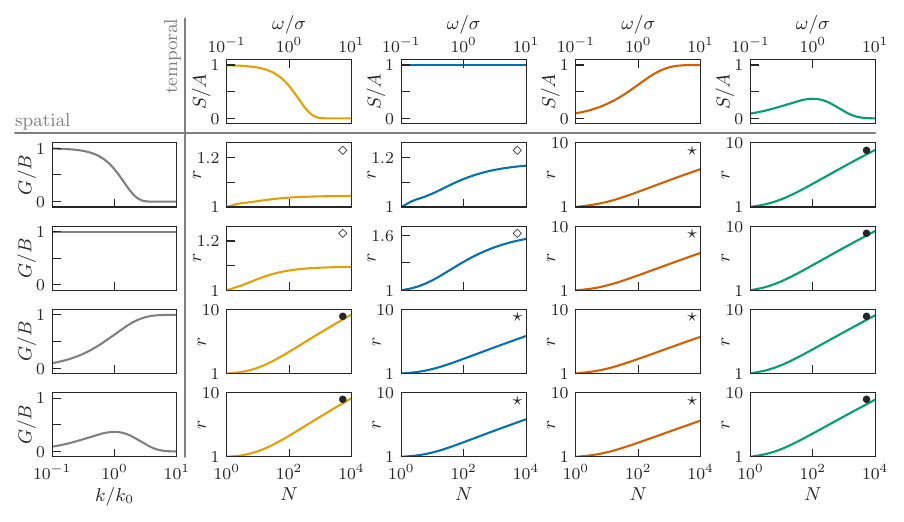}
  \caption{The metrological gain $r$ of estimation uncertainties for separable and spin-squeezed protocols for different spatial and temporal noise correlations. The first row and first column of the grid show the considered temporal $S(\omega)$ and spatial noise $G(k)$ spectra, respectively, which are combined in the inner plots. From top to bottom and left to right, the spectra are the following: Gaussian, white, linear, and Ohmic. The temporal (spatial) spectra are characterized by strength $A$ ($B$) and width $\sigma$ ($k_0$), with $A/\sigma = 1/8$ ($B / k_0 = 8$). Each inner plot displays $r$ as a function of the number of qubits $N$ on a log-log scale, except for ($\diamond$) where the y-axis is scaled linearly. Symbols indicate the scaling behavior of $r$: ($\diamond$) marks cases where the estimation uncertainty only surpasses the standard quantum limit by a constant factor, ($\bullet$) corresponds to $r \propto N^{1/4}$, and ($\star$) to $r \propto N^{1/6}$.}
  \label{fig:fig4}
\end{figure*}
Finally, we discuss general spatial and temporal noise, where we assume that the spatial and temporal correlations can be separated,
\begin{equation}
\label{eq:spatial_cor_noise}
    \Ex{\phi_{nl\tau}\phi_{n'l'\tau}} = Q_\tau(l-l') P(n-n').
\end{equation}
In particular we define $\gamma(\tau) = Q_\tau(0) P(0)$. In the temporal case, $Q_\tau(l-l')$ describes the correlation of the temporally integrated noise. Spatially, however, we assume the probes to have no spatial extension. Hence, $P(n-n')$ represents the correlation between discrete points. We consider the probes to be aligned in a line with fixed spacing $\Delta_x$. The Nyquist frequency in this setting is given by half the spatial sampling rate, $\pi/\Delta_x$. Accordingly, we define the spatial noise via its power spectrum $G(k)$, with $k \in [-\pi/\Delta_x, \pi/\Delta_x]$. By the Wiener-Khinchin theorem~\cite{Ziemer2014}, the spatial correlation function $P(n-n')$, which represents the spatial part of the correlation function $\Ex{F_n(0) F_{n'}(0)}$, is the inverse Fourier transform of the power spectrum. For simplicity, we choose $\Delta_x = 1$, leading to:
\begin{equation}
    P(n-n') = \frac{1}{2\pi} \int_{-\pi}^\pi G(k) \eexp{i k (n - n')} dk.
\end{equation}

We discuss the same noise spectra as for temporal noise in the previous section [see Eq.~\eqref{eq:spectrum_tempnoise}], but restricted to the range $k \in [-\pi, \pi]$. The estimation uncertainty in this scenario is described by Eq.~\eqref{eq:uncertainty_total}. To analyze the expression, we expand $\sinh{x}$ and $\cosh{x}$ to lowest order in $x$, neglect terms of order $1/N^2$, take the long time limit, and assume the spatial noise correlation length to be much shorter than the number of particles. This leads to
\begin{equation}
\begin{aligned}
\label{eq:uncertainty_temp_cor}
   \Var{\est} &\approx \frac{\Ex{{J_y}^2(0)}}{\tau T \Ex{J_z(0)}^2} + \frac{N \left[\eexp{\gamma(\tau)} - 1\right]}{4 \tau T \Ex{J_z(0)}^2} - \frac{\gamma(\tau)}{N \tau T}\\
   &+ \frac{\gamma(\tau)}{N \tau T} \frac{\left[ \Delta J_z(0) \right]^2}{\Ex{J_z(0)}^2} \! \left[ \frac{G(0)}{P(0)} \! - \!1 \right] \! + \! \frac{S(0)G(0)}{N T}.
\end{aligned}
\end{equation}

Comparing this equation with the simplified expression for spatially uncorrelated noise, we recover Eq.~\eqref{eq:uncertainty_temp_simp} by setting $G(0) = P(0) = 1$, and we see that the first three terms remain unchanged. The last term, resembling the fundamental limit of Eq.~\eqref{eq:fund_spatial}, is decisive for whether a scaling beyond $\Delta \est \propto 1/\sqrt{N}$ is achievable. While in the case of spatial white noise, an improved scaling required $S(0) = 0$, in the presence of spatial noise correlations, either $S(0) = 0$ or $G(0) = 0$ is sufficient. The fourth term of Eq.~\eqref{eq:uncertainty_temp_cor} grows with increasing squeezing, as $\Delta J_z(0)$ is zero for separable states and increases with decreasing $\kappa$. Its sign depends on whether the spatial noise is correlated or anti-correlated, $[G(0)/P(0) - 1] \geq -1$. To leading order, the fourth term scales with $\kappa$ and $\tau$ the same way as the sum of the second and third terms, where the sum of the three terms is positive. Moreover, the scaling in $\tau$ of $\gamma(\tau)$ is not influenced by the spatial noise spectrum. As a result, minimizing Eq.~\eqref{eq:uncertainty_temp_cor} to leading order results in the same $N$ scaling of the optimal parameters $\tau_{\text{opt}}$ and $\kappa_{\text{opt}}$ as reported in Tab.~\ref{tab:noise-scaling} for spatially uncorrelated noise. 

Fig.~\ref{fig:fig4} shows the dependence of the gain $r$ on the number of qubits $N$ for different combinations of spatial and temporal noise. We see that for the cases where neither $G(0) = 0$ nor $S(0) = 0$, i.e., the four combinations of Gaussian and white spectra, squeezing improves the uncertainty only up to a constant factor compared to the separable case. In these cases, we find that the advantage of using squeezing decreases with stronger noise correlations. Hence, the weakest entanglement enhancement of the measurement uncertainty is observed for noise with both temporal and spatial Gaussian correlations. Among the noise spectra in Fig.~\ref{fig:fig4} with $S(0) = 0$ or $G(0) = 0$, we observe the best scaling of the metrological gain $r$ for Gaussian and Ohmic temporal noise. In both cases, we have $\gamma(\tau) \propto \tau^2$, resulting in $r \propto N^{1/4}$. For linear and white temporal noise, $\gamma(\tau)$ only decreases linearly in $\tau$ with shorter interrogation times, leading to $r \propto N^{1/6}$.

These results illustrate that temporal and spatial noise correlations are not metrologically identical. They play a similar role in the question of whether a scaling above the standard quantum limit is possible. However, the exact scaling is determined by the shape of the temporal spectrum.

\section{Conclusion}
In this work, we have demonstrated that quantum entanglement can provide a considerable gain in the sensitivity of quantum sensing compared to Ramsey spectroscopy with separable states. Crucially, our analysis extends previous work~\cite{Matsuzaki2011, Chin2012} by including correlations between successive experimental shots, which arise in many practical scenarios where the preparation of the system between shots does not reset the noise. Our analysis shows that an uncertainty scaling beyond the standard quantum limit of $1/\sqrt{N}$ can be achieved for classical pure dephasing noise if the zero-frequency component of the noise spectrum is zero. Intuitively speaking, this noise component at zero frequency is indistinguishable from the signal itself and therefore provides a lower limit to the uncertainty set by the signal-to-noise ratio of the signal observed by the sensors. For noise with a vanishing zero-frequency component, the optimal scaling using spin-squeezed states depends on the short-time behavior of the dephasing $\gamma(\tau)$. In the case of $\gamma(\tau) \propto \tau$, where there is no advantage to expect from GHZ states over separable states, the use of spin-squeezed states enables an uncertainty scaling as $\Delta \est \propto N^{-2/3}$. For $\gamma(\tau) \propto \tau^2$, we reach an optimal scaling of $\Delta \est \propto N^{-3/4}$, which corresponds to the scaling reported for GHZ states in Refs.~\cite{Matsuzaki2011, Chin2012}. However, these works considered noise that was uncorrelated between shots of the experiments and are hence not directly comparable. While this scaling is proven to be optimal in the case of no correlations among different Ramsey sequences~\cite{Macieszczak2015}, its optimality in the presence of correlated experimental shots remains an open question. Furthermore, we emphasize that we only consider a particular sensing strategy based on standard Ramsey spectroscopy. For the considered noise sources, it may be possible to devise more optimal strategies both with and without entanglement.

\section*{Acknowledgments}

The authors thank Klaus Mølmer, Susana Huelga, and Martin Plenio for helpful comments. This work was supported by the Novo Nordisk Foundation (Challenge
project “Solid-Q”) and Danmarks Grundforskningsfond (Grant No. DNRF139, Hy-Q Center for Hybrid Quantum Networks).

\appendix

\section{Spin-squeezed states}
\label{app:spinsqueezing}

\begin{figure}[]
  \centering
  \includegraphics[width=\columnwidth]{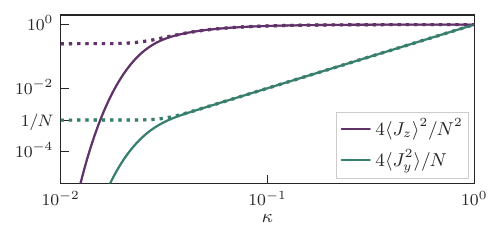}
\caption{Dependence of the expectation values most relevant to the estimation uncertainty on the squeezing parameter $\kappa$ for $N = 1000$ (solid lines) and
$N = 1001$ (dashed lines) qubits. The different behavior as $\kappa \rightarrow 0$ occurs because states with an even number of qubits approach $\ket{\psi_{\kappa\rightarrow0}} = \ket{0}_y$, while states with an odd number of qubits converge to $\ket{\psi_{\kappa\rightarrow0}} \propto \ket{1/2}_y - \ket{-1/2}_y$.}
  \label{fig:fig5}
\end{figure}

In this appendix, we extend the discussion on squeezing enhanced sensitivity of Ramsey spectroscopy to other classes of spin-squeezed states, particularly those considered more practical. So far, we focused on the spin squeezed states $\ket{\psi_\kappa}$ defined in Eq.~\eqref{eq:squeezed_states}. The motivation for this choice is that their spin expectation values can be obtained easily, and varying $\kappa$ allows one to continuously tune the entanglement from an approximately coherent state at $\kappa = 1$ to a maximally squeezed state with $\Delta J_y / \Ex{J_z} \propto 1/N$ for $\kappa \rightarrow 0$, as illustrated in Fig.~\ref{fig:fig5}. Other prominent classes of spin-squeezed states are one-axis twisted (OAT) and two-axis twisted (TAT) states~\cite{Kitagawa1993}. From a practical perspective, these states are relevant, as they can be prepared by evolving a coherent spin state with the nonlinear Hamiltonians 
\begin{equation}
    H_{\text{OAT}} = \chi J_z^2, \quad H_{\text{TAT}} = \frac{\chi}{2 i} \left( J_+^2 - J_-^2 \right),
\end{equation}
where $J_\pm$ are the collective spin ladder operators and $\chi$ is a constant~\cite{Kitagawa1993}. OATs have been realized experimentally in several hardware platforms through collective nonlinear interactions~\cite{Pezze2018}. Recently, two-axis twisting has been demonstrated with rubidium atoms~\cite{Luo2025}.

\begin{figure}[]
\centering
\includegraphics[width=\columnwidth]{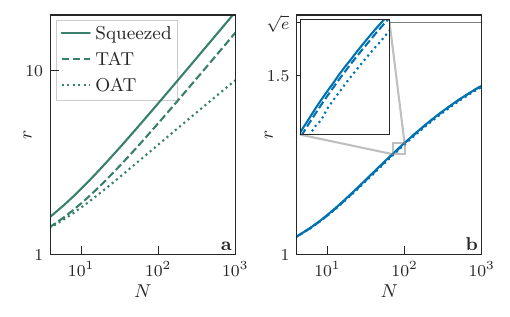}
  \caption{Comparison of the metrological gain $r$ in Ramsey spectroscopy between spin-squeezed states $\ket{\psi_\kappa}$, defined in Eq.~\eqref{eq:squeezed_states}, one-axis twisted states (OAT), and two-axis twisted states (TAT), as a function of the number of probes $N$. The metrological gain is calculated relative to standard Ramsey spectroscopy with separable input states and is shown for the cases of (a) no noise and (b) white noise.}
  \label{fig:fig6}
\end{figure}

Fig.~\ref{fig:fig6} compares the metrological gain of the three states over standard Ramsey spectroscopy with separable input states for uncorrelated noise and in the absence of noise. The figure is obtained by evaluating Eq.~\eqref{eq:uncertainty_nonoise} and Eq.~\eqref{eq:uncertainty_markovian}, for even values of $N$ and optimizing the interrogation time and squeezing, where $\Ex{J_z}$ and $\Delta{J_y}$ are analytically evaluated for OATs~\cite{Ma2011}, and numerically evaluated for $\ket{\psi_\kappa}$ and TATs. In the noise-free case, Fig.~\hyperref[fig:fig6]{\ref{fig:fig6}a}, the metrological gain is proportional to $\Ex{J_z} / (\sqrt{N}\,\Delta J_y)$ [see Eq.~\eqref{eq:uncertainty_nonoise}]. In the limit of large $N$, $\Delta J_y / \Ex{J_z} \propto 1 / N$ for TATs and $\ket{\psi_\kappa}$, leading to $r \propto \sqrt{N}$~\cite{Kitagawa1993}. OATs reach $\Delta J_y / \Ex{J_z} \propto N^{-5/6}$, resulting in $r \propto \sqrt[3]{N}$~\cite{Kitagawa1993}. For white noise, Fig.~\hyperref[fig:fig6]{\ref{fig:fig6}b}, the relative ordering of the three states in terms of metrological gain remains the same. As discussed in Sec.~\ref{subsec:markoviannoise} and displayed in Fig.~\hyperref[fig:fig2]{\ref{fig:fig2}b}, the metrological gain is upper-bounded by $\sqrt{e}$. The observed differences are much smaller than in the noise-free case. We verified that all three classes of squeezed states exhibit similar behavior under the colored noise considered in Fig.~\ref{fig:fig3}, although OATs do not achieve the same metrological advantage as the other two classes of states for Ohmic noise, since OAT is not able to produce sufficient amounts of squeezing. 

\section{Noise properties}
\label{app:dephasingrate}
In this appendix, we provide the exact expressions for the dephasing behavior of the temporal noise spectra used in Sec.~\ref{sec:uncertainty_analysis}. Tab.~\ref{tab:noise-scaling} shows the short- and long-time limits of these expressions to the leading order. Additionally, we present the autocorrelation functions of these spectra to illustrate their temporal correlations and anti-correlations.

The analysis of the simplified uncertainty expressions of Eq.~\eqref{eq:uncertainty_temp_simp} and Eq.~\eqref{eq:uncertainty_temp_cor} shows the influence of the dephasing $\gamma(\tau) = \Ex{\phi_{nl\tau^2}}$ on the uncertainty scaling. To evaluate $\gamma(\tau)$, we set $\Delta_l = 0$ in Eq.~\eqref{eq:formula_noise_correlations} and use the noise spectra defined in Eq.~\eqref{eq:spectrum_tempnoise}. This results in the following dephasing behavior:
\begin{equation}
    \begin{aligned}
        \gamma_W(\tau) \!= & A \tau \\
        \gamma_G(\tau) \!= & A \tau \! \left[ \frac{1}{\tau \sigma} \sqrt{\frac{2}{\pi}} \left( \eexp{-\frac{\sigma^2 \tau^2}{2}} - 1 \right) + \text{erf}\left( \frac{\tau \sigma}{\sqrt{2}}\right) \! \right]\\
        \gamma_L(\tau) \!= & A \tau \!\left[\! 1  \! - \! \frac{2}{\pi} \! \arctan(\sigma \tau) \! + \! \frac{1}{\pi \sigma \tau} \log\!\left(1 + \sigma^2 \! \tau^2\right)\!\right]\\
        \gamma_O(\tau) \!= & A\tau \! \left[ \frac{1}{\pi \sigma \tau} \log\left( 1 + \sigma^2 \tau^2 \right) \! \right],
    \end{aligned}
\end{equation}
where the subscripts $W$, $G$, $L$, and $O$ refer to the white, Gaussian, linear, and Ohmic noise spectra as defined in Eq.~\eqref{eq:spectrum_tempnoise}. These functions are plotted in Fig.~\hyperref[fig:fig7]{\ref{fig:fig7}b}. Their short- and long-time behavior, summarized in Tab.~\ref{tab:noise-scaling}, is obtained by approximating the involved functions in the regimes $\sigma \tau \ll 1$ and $\sigma \tau \gg 1$.

According to the Wiener-Khinchin theorem, the noise spectra $S(\omega)$ and the autocorrelation function $R(t)$ are related by a Fourier transform~\cite{Ziemer2014}. Using the non-unitary Fourier convention, this relation reads:
\begin{equation}
    R(t) = \frac{1}{2 \pi}\int_{-\infty}^\infty S(\omega) \eexp{i \omega t} d\omega.
\end{equation}
Evaluating this relation for the noise spectra in Eq.~\eqref{eq:spectrum_tempnoise} gives the corresponding autocorrelation functions:
\begin{equation}
    \begin{aligned}
        R_W(t) \!= & A \delta(t)\\
        R_G(t) \!= & \frac{A \sigma}{\sqrt{2 \pi}} \eexp{-\frac{\sigma^2 t^2}{2}} \\
        R_L(t) \!= & A \left[ \delta(t) - \frac{\sigma}{\pi} \frac{1}{1 + \sigma^2 t^2}\right] \\
        R_O(t) \!= & \frac{A \sigma}{\pi} \frac{1-\sigma^2 t^2}{\left(1 + \sigma^2 t^2\right)^2},
    \end{aligned}
\end{equation}
where the subscripts $W$, $G$, $L$, and $O$ refer to white, Gaussian, linear, and Ohmic noise spectra as defined in Eq.~\eqref{eq:spectrum_tempnoise}. In particular, we see that Ohmic noise is correlated for $t \, \sigma < 1$ and anti-correlated for $t \, \sigma > 1$. 

\section{Time dependence}
\label{app:timediscussion}
In this appendix, we formalize the notion of taking the long-time limit. Furthermore, we briefly discuss the short-time behavior in the case of a single qubit for all of the considered noise spectra.

In the uncertainty analysis of Sec.~\ref{sec:uncertainty_analysis}, we considered the long-time limit of Eq.~\eqref{eq:uncertainty_total}. This is reached when the third term on the right-hand side of Eq.~\eqref{eq:uncertainty_total} becomes proportional to $L = T/\tau$, such that all terms in the numerator scale linearly in $T$. Combined with the $T^2$ in the denominator, this leads to an overall scaling as $\sqrt{1/T}$. To achieve this, we require that
\begin{equation}
    \sum_{l \neq l'}^L \sinh{\left(\Ex{\phi_{nl\tau}\phi_{n'l'\tau}}\right)} \propto L,
\end{equation}
which holds when $T$ is much longer than the correlation time of the noise. In this regime, we can use the relation 
 \begin{equation}
     \lim_{T \to \infty} \frac{\gamma(T)}{T} = S(0),
 \end{equation}
to simplify Eq.~\eqref{eq:uncertainty_temp_simp} and recover the fundamental limit of Eq.~\eqref{eq:fund_spatial}. 

\begin{figure}[t]
  \centering
\includegraphics[width=\columnwidth]{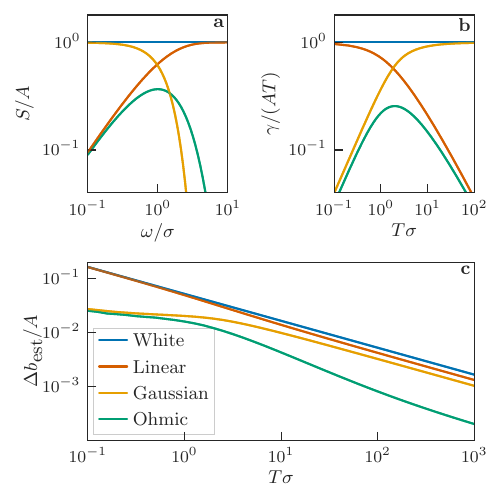}
  \caption{Estimation uncertainty of Ramsey spectroscopy with a single particle as a function of the total time $T$ for different noise spectra.  (a) Plot of the considered white, Gaussian, linear, and Ohmic noise spectra $S(\omega)$, with $\sigma / A = 1/10^3$. (b) Normalized dephasing rate $\gamma(T)/T$ as a function of the total time $T$. For white noise $\gamma(T) \propto T$, Gaussian noise with temporal correlations leads to $\gamma(T) \propto T^2$ for $T\sigma <1$, the anti-correlations of linear noise result in noise cancellations, and for Ohmic noise, the noise is first correlated and for longer times anti-correlated, resulting in noise $\gamma(t)/(AT)$ peaking around $\sigma T=1$. (c) Corresponding single qubit estimation uncertainty $\Delta \est$ as a function of $T$.}
  \label{fig:fig7}
\end{figure}

For shorter times $T$, the situation is more complex. Fig.~\ref{fig:fig7} shows the dependence of the estimation uncertainty $\Delta \est$ on the total time $T$ for Ramsey spectroscopy with a single particle. In the case of white noise, the noise is not correlated between the experiments, and the uncertainty scales as $1/\sqrt{T}$ for all $T$. For linear noise, anti-correlations accumulate, visible in Fig.~\hyperref[fig:fig7]{\ref{fig:fig7}c} at $\sigma T \approx 1$. For $T \sigma \ll 1$, $\gamma(T)/T \propto 1$, while for $T \sigma \gg 1$, $\gamma(T)/T \propto \log(T)/T$. Although this does not affect the overall scaling, it reduces the prefactor around $T \sigma \approx 1$. When the noise spectrum is Gaussian, for $T \sigma \ll 1$ all experiments are exposed to strongly correlated dephasing $\Ex{\phi_{nl\tau}\phi_{n'l'\tau}} \approx \delta_{n,n'} \gamma(\tau)$. In this regime, additional repetitions do not average out the noise, giving $\gamma(T)/T \propto T$, resulting in almost no improvement in the uncertainty with increasing $T$. For $T\sigma \gg 1$, most of the experiments are uncorrelated, so that $\gamma(T)/T \propto 1$, and the standard $1/\sqrt{T}$ scaling is recovered. For noise with an Ohmic spectrum, noise correlations, similar to the Gaussian case, for $T \sigma \ll 1$ are combined with the properties of linear noise in the regime $T \sigma \gg 1$. This leads to a scaling almost independent of $T$ for $T \sigma \ll 1$ and a scaling $1/\sqrt{T}$ for $T \sigma \gg 1$.

\section{GHZ states}
\label{app:GHZ}

In the discussion of entanglement-enhanced Ramsey spectroscopy, GHZ states, alongside spin-squeezed states, are typically considered. Different from spin-squeezed states, where the enhancement originates from the lower shot noise, GHZ states improve the sensitivity as the phase of a collective  $N$-qubit state evolves $N$ times faster compared to a single qubit state~\cite{Orgikh2001}. In the absence of noise, GHZ states are optimal and achieve Heisenberg scaling~\cite{Bollinger1996}. Moreover, they were used in Refs.~\cite{Matsuzaki2011} and \cite{Chin2012} to demonstrate an enhanced uncertainty scaling in the presence of non-Markovian dephasing noise that is uncorrelated between subsequent experiments. In this section, we repeat the derivation from Sec.~\ref{sec:uncertainty_derivation} using GHZ states as input. For simplicity, we restrict the derivation of the estimation uncertainty and its analysis to spatially uncorrelated noise. 

The protocol for Ramsey spectroscopy with GHZ states consists of the following steps:
\begin{enumerate}[noitemsep]
    \item Preparation of an $N$-qubit GHZ state $\ket{\text{GHZ}} = \frac{1}{\sqrt{2}}\left(\ket{0}^{\otimes N} + \ket{1}^{\otimes N}\right)$;
    \item Free evolution of the system for a time $\tau$;
    \item Projection measurement onto the initial GHZ state;
    \item Repetition of steps 1-3 for $L = T/\tau$ iterations.
\end{enumerate}
Unlike the protocols for separable input states and spin-squeezed input states, this protocol does not involve Ramsey pulses. The parameters of the protocol to optimize are the operating point $\sig_r \tau$ and the interrogation time $\tau$. 

Given the combined state of all $L$ experimental runs and $N$ qubits, we define the measurement operator for the $l$-th experiment that operates on this total state as
\begin{equation}
    M_l(0) \! = \! 2 \! \left( \bigotimes_{i=1}^{l-1} \mathbb{I}_i \! \otimes \! \ket{\text{GHZ}}_l \! \bra{\text{GHZ}} \! \otimes \! \bigotimes_{i=l+1}^L \mathbb{I}_i\right) \! - \mathbb{I}^{\otimes L},
\end{equation}
where we have rescaled the measurement operator such that the measurement outcomes are in $[-1, 1]$. Following the derivation in Sec.~\ref{sec:uncertainty_derivation}, we find $M_l(\tau)$ in the Heisenberg picture with the effective Hamiltonian of Eq.~\eqref{eq:H_eff} for a fixed interrogation time $\tau$. The evolution of the parts proportional to the identity is trivial. The GHZ state $\ket{\text{GHZ}}_l$ in the projection operator evolves as:
\begin{equation}
\begin{aligned}
    \ket{\text{GHZ}(\tau)}_l &= \frac{1}{\sqrt{2}}\left[\ket{0}^{\otimes N} \! + \! e^{i \Sigma_n \left(\sig_r\tau + \phi_{nl\tau}\right)} \ket{1}^{\otimes N}\right] \\
    &= \frac{1}{\sqrt{2}}\left[\ket{0}^{\otimes N} \! + \! e^{i \left(\theta + \Sigma_n \, \phi_{nl\tau} \right)} \ket{1}^{\otimes N}\right],
\end{aligned}
\end{equation}
where we have introduced the phase $\theta = N \sig_r \tau$. The expectation value of the $l$-th measurement is obtained by computing the inner product between the evolved GHZ state and the initial GHZ state, followed by averaging over noise trajectories. Assuming spatially uncorrelated, Gaussian-distributed noise, we find
\begin{equation}
    \Ex{M_l(\tau)} = e^{-\frac{N \gamma(\tau)}{2}} \cos(\theta).
\end{equation}
The response of the measurement to signal changes is given by
\begin{equation}
    \abs{\frac{\partial\Ex{M_l(\tau)}}{\partial \theta} \frac{\partial\theta}{\partial \sig}} = N \tau e^{-\frac{N \gamma(\tau)}{2}} \abs{\sin(\theta)},
\end{equation}
which reaches its maximum at $\theta = q\pi/2$ for odd integers $q$. Choosing this value as the operating point of the sensor, we can evaluate the estimation uncertainty,
\begin{equation}
    \Delta \est = \frac{\Delta \left[ \sum_l M_l(\tau) \right]}{L \abs{\frac{\partial\Ex{M_0(\tau)}}{\partial \sig}}}.
\end{equation}
Applying the same identities introduced in Sec.~\ref{sec:uncertainty_derivation}, we arrive at the following expression for the estimation uncertainty in Ramsey spectroscopy with GHZ states:
\begin{equation}
    \left(\Delta \est\right)^2 = \frac{L e^{N \gamma(\tau)} + \sum_{l \neq l'} \sinh\left( N \Ex{\phi_{0l\tau}\phi_{0l'\tau}} \right)}{N^2 T^2}. 
\end{equation}

In the absence of noise, we recover the Heisenberg scaling $\Delta \est = 1 / \left(N \sqrt{T\tau} \right)$~\cite{Bollinger1996}. For Markovian noise $\gamma(\tau) = A\tau$, the optimal interrogation time is $\tau = 1/(NA)$, leading to an uncertainty of $\Delta \est = \sqrt{Ae}/\sqrt{NT}$. This matches the estimation uncertainty obtained with separable input states~\cite{Huelga1997}, resulting in no metrological gain $r=1$ independent of $N$. 

\begin{figure}[t]
\centering
\includegraphics[width=\columnwidth]{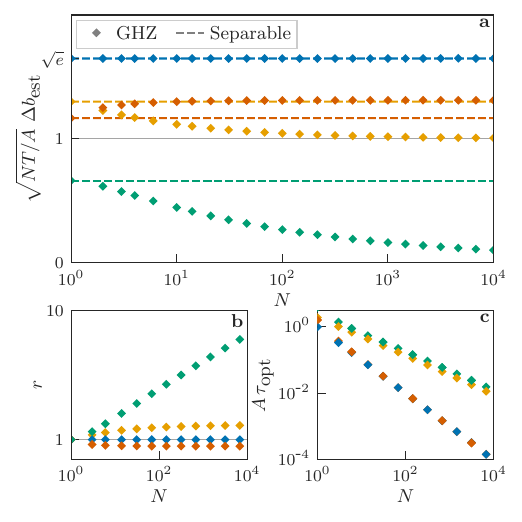}
  \caption{Estimation uncertainty in Ramsey spectroscopy with GHZ input states under various noise spectra (plotted in Fig.~\hyperref[fig:fig3]{\ref{fig:fig3}a}), for a certain total probing time $T$ much larger than the noise correlation time and dephasing time. (a) Scaled uncertainty $\sqrt{NT / A} \, \, \Delta \est$ as a function of the number of qubits $N$ for GHZ input states compared to separable input states with (counting from below at $N=1$) Ohmic, linear, Gaussian, and white spectra. (b) Metrological gain $r$ of estimation uncertainties between separable and GHZ input states as a function of $N$. (c) Optimal interrogation time $\tau_{\text{opt}}$ for GHZ input states versus $N$.
}
\label{fig:fig8}
\end{figure}

Fig.~\ref{fig:fig8} shows the result of estimation using GHZ states. It is the equivalent of Fig.~\ref{fig:fig3} but for GHZ- instead of spin-squeezed states. In Fig.~\hyperref[fig:fig8]{\ref{fig:fig8}a}, the minimal normalized uncertainty is plotted against the number of qubits. We observe the above-discussed absence of an advantage of GHZ-states over separable states in the case of white noise, leading to $r=1$ in Fig.~\hyperref[fig:fig8]{\ref{fig:fig8}b}.

For noise with Gaussian and Ohmic spectra, where $\gamma(\tau) \propto \tau^2$ at short interrogation times, the optimal interrogation time asymptotically scales as $\tau_{\text{opt}} \propto N^{-3/5}$, see Fig.~\hyperref[fig:fig8]{\ref{fig:fig8}c}. For Ohmic noise, this results in an estimation uncertainty scaling of $\Delta \est \propto N^{-7/10}$, shown in Fig.~\hyperref[fig:fig8]{\ref{fig:fig8}a}, and hence a metrological gain $r \propto N^{1/5}$, shown in Fig.~\hyperref[fig:fig8]{\ref{fig:fig8}b}. This scaling advantage is slightly weaker than the $r\propto N^{1/4}$ achieved with spin-squeezed states. Gaussian noise results in $\Delta \est \propto 1/\sqrt{N}$, as the fundamental limit of Eq.~\eqref{eq:fund_spatial} still applies. 

A difference in the scaling of the estimation uncertainty between GHZ states and spin-squeezed states appears for noise with a linear spectrum. While spin-squeezed input states offer a scaling advantage over separable states, GHZ states perform worse than separable input states. For the linear spectrum [see Eq.~\eqref{eq:spectrum_tempnoise}] the dephasing $\gamma(\tau)$ can in the regime $\sigma \tau < 1$ be approximated as $\gamma(\tau) \approx A \tau \left(1 - \frac{\tau \sigma}{\pi} \right)$. When $\sigma \tau / \pi < 1$, the linear behavior dominates, and we write $\gamma(\tau)\approx \tilde{A} \tau$, where the noise strength $\tilde{A}$ is weakly time dependent. This leads to an estimation uncertainty $\Delta \est = [\tilde{A}(e - 1)/(NT)]^{1/2}$ for both GHZ and separable input states. However, while for GHZ states $\tilde{A}$ approaches $A$ for large $N$, $\tilde{A}$ is smaller for separable states due to the longer optimal interrogation time. Consequently, separable states achieve a lower estimation uncertainty than GHZ states, as shown in Fig.~\ref{fig:fig8}.

Spin-squeezed states have the advantage that the amount of entanglement is tunable. This freedom allows one to achieve better performance with spin-squeezed states than with separable or maximally entangled states. A similar advantage could perhaps be gained with GHZ states by preparing a state $\ket{\Phi}$ consisting of $N/S$ GHZ states of size $S$, 

\begin{equation}
    \ket{\Phi} = \bigotimes_{i=1}^{N/S} \frac{1}{\sqrt{2}} \, \left(\ket{0}^{\otimes S} + \ket{1}^{\otimes S}\right).
\end{equation}

\noindent For spatially uncorrelated noise, the measurement outcomes for each of the $N/S$ blocks are uncorrelated. For fixed size $S$, this leads to a total estimation uncertainty scaling proportional to $1 / \sqrt{N}$, due to the statistical averaging over $N/S$ independent estimates. Hence, for spatially uncorrelated noise, this strategy does not provide a scaling advantage over separable input states. However, the situation may change when the noise is spatially correlated, opening up many different strategies to exploit the spatial structure. A detailed discussion of this possibility is, however, beyond the scope of this work. 

\bibliography{References}

\end{document}